\documentclass[twocolumn,superscriptaddress,pre]{revtex4}

\usepackage{amsmath}
\usepackage{amsbsy}
\usepackage{bbm}
\usepackage{dcolumn}
\usepackage{enumitem}
\usepackage{graphicx}
\usepackage{float}

\newcommand\argmax{\mathop\mathrm{argmax}\limits}
\newcommand\bm{\boldsymbol}
\newcommand\etal{\textit{et al.}}

\begin{document}
\title{Mixing patterns and individual differences in networks}
\author{George T. Cantwell}
\affiliation{Department of Physics, University of Michigan, Ann Arbor, Michigan, USA}
\author{M. E. J. Newman}
\affiliation{Department of Physics, University of Michigan, Ann Arbor, Michigan, USA}
\affiliation{Center for the Study of Complex Systems, University of Michigan, Ann Arbor, Michigan, USA}

\begin{abstract}
  We study mixing patterns in networks, meaning the propensity for nodes of different kinds to connect to one another.  The phenomenon of assortative mixing, whereby nodes prefer to connect to others that are similar to themselves, has been widely studied, but here we go further and examine how and to what extent nodes that are otherwise similar can have different preferences.  Many individuals in a friendship network, for instance, may prefer friends who are roughly the same age as themselves, but some may display a preference for older or younger friends.  We introduce a network model that captures this behavior and a method for fitting it to empirical network data.  We propose metrics to characterize the mean and variation of mixing patterns and show how to infer their values from the fitted model, either using maximum-likelihood estimates of model parameters or in a Bayesian framework that does not require fixing any parameters.
\end{abstract}

\maketitle

\section{Introduction}
Networks are widely used to represent patterns of connections in complex systems, such as the Internet, the World Wide Web, and social and biological networks.  A~common feature of many networks is \textit{assortative mixing}, the tendency of network nodes to be connected to others that are similar to themselves in some way~\cite{MSC01,Newman03c,BG14, Noldus15}.  
On the World Wide Web, for instance, one might expect web pages to link to others written in the same language more than they do to ones in different languages.  In friendship networks (where the phenomenon is also known as \textit{homophily}) many individuals have a preference for friends who are similar to themselves in terms of age, race, educational level, and other characteristics~\cite{MSC01,Moody01,Newman03c,BG14}.  One can also encounter \textit{disassortative} mixing, the tendency for nodes to connect to unlike others~\cite{Newman03c, Noldus15}.

Assortative mixing has been studied widely.  Researchers have examined and quantified assortativity as it occurs in a wide variety of real-world networks~\cite{MSC01,Newman03c,BG14} and created mathematical models such as the planted partition model~\cite{CK01,McSherry01} and the stochastic block model~\cite{HLL83} that can mimic both assortative and disassortative behaviors.  These methods and models, however, capture only the average mixing behavior of nodes, the average preference for members of one group to forge connections with another.  There can be, and in many cases is, substantial variation about the average; all members of a group do not necessarily behave the same.

As an example, networks of romantic interaction between individuals, which are widely studied in sociology, are mostly disassortative by gender: a majority of individuals have a preference for romantic engagements with members of the opposite sex.  On the other hand, some people prefer romantic engagements with the same sex. Standard measures of overall assortative mixing would thus say that the average individual has a small fraction of same-sex relationships and the rest are opposite-sex.  But this is misleading: in fact, many individuals have strong preferences for one or the other, so the ``average preference'' does not, in this case, provide a good description of individual behaviors.

Furthermore, there can be interesting mixing patterns even when there is little or no average assortativity in a network.  For example, a recent study of friend networks on Facebook showed little to no gender assortativity on average, yet \emph{some} people do appear to have preferences~\cite{ugander2011anatomy,altenburger2018monophily}.  Some individuals on Facebook strongly prefer either male or female friends---it is only when we average over the whole population that we see no effect.  Thus, traditional measures of average assortativity do not tell the whole story.

There has been some previous literature discussing these phenomena and advocating a move beyond average measures of assortativity.  In the study of Facebook mentioned above, Altenburger and Ugander~\cite{altenburger2018monophily} introduced the concept of \textit{monophily}, the extent to which people's friends are similar to one another, while Peel~\etal~\cite{peel2018multiscale} define a variant assortativity coefficient that characterizes assortativity within a local neighborhood in a network.  Other approaches have defined an assortativity coefficient at the level of individual nodes~\cite{Piraveenan08, Piraveenan12}.

In this paper we demonstrate that inferring and quantifying individual differences in mixing is not trivial, in practice or in principle, an observation that bears emphasizing.  The difficulty is not simply due to a lack of data.  Even for arbitrarily large networks naive approaches will fail.  To address these issues we introduce a principled and general method for analyzing mixing patterns in networks that does not require large amounts of data or lengthy computations.  Our solution employs a generative stochastic model of individual-level mixing, showing how it can be used to model and analyze empirical network data.  Crucially, the model allows for arbitrary mixing patterns and does not assume that individuals behave in accordance with the average within their group.  By fitting the model to data using statistical methods we infer quantities that have straightforward interpretations and can thus be used to characterize mixing patterns, in much the same way that the parameters of a normal distribution characterize mean and variance.

The model we study is conceptually similar to others that have been studied previously.  It shares with the well-known stochastic block model~\cite{HLL83} the ability to represent arbitrary mixing patterns at the group level, but also goes further, allowing for individual variation within the groups.  A model for individual variation was introduced previously in~\cite{altenburger2018monophily}, but it does not allow for arbitrary mixing patterns, nor was a direct method proposed to fit the model to data.  Variation within groups can be approximated with mixed membership models~\cite{Airoldi08,Latouche11}, in which network nodes can be members of multiple groups and inherit the mixing patterns of all of their groups.  Two nodes in a given group might, for instance, be in different other groups and hence need not mix equivalently.  This approach is of little use, however, when group memberships are already known or the categories are known to be distinct and non-overlapping.  If we want to model individual differences in the social mixing patterns of men and women, for instance, we are not at liberty to re-assign genders so that our model fits.

As a demonstration of our methods, we apply them to two example networks, a friendship network of high school students and a linguistic network of word adjacencies in English text.  We find that there is indeed substantial individual variation in mixing patterns in both networks, implying that traditional average measures of mixing offer an incomplete description of network structure.

\section{Individual preferences and patterns of connection}
\label{sec:preferences}
We consider networks in which the nodes are divided into a number of discrete, non-overlapping groups, types, or categories, and where individual nodes have \textit{preferences} about the types of the nodes with which they have network connections.  We will focus on labeled networks, meaning ones in which the type of every node is known in advance---we are told the sex of each individual in a social network, for example, or the language that each web page is written in.  Our network could be directed or undirected, but we will concentrate primarily on the directed case here, treating the undirected one as the special case when all edges are reciprocated.

In the context of such labeled network data, how should one define preference? By any reasonable definition, if a node has a strong preference to connect to others of a certain type then we should expect there to be a relatively large number of edges to that type.  Let us denote the number of edges from node~$i$ to nodes of type~$s$ by~$k_{is}$ and the total number of edges from~$i$ to nodes of all types by $k_i = \sum_s k_{is}$.  Then the ratio $k_{i s}/ k_i$ is the fraction of edges from node~$i$ to nodes of type~$s$.

This ratio, however, is not necessarily an accurate guide to $i$'s preference for connections to type~$s$.  We should expect there to be some statistical fluctuations in the network formation process, so that high or low values of $k_{is}$ could occur just by chance.  Let us define a quantity~$x_{is}$ to represent $i$'s underlying preference for nodes of type~$s$, which will be equal to the expected value of the ratio $k_{is}/k_i$, averaged over these fluctuations:
\begin{equation}
  \label{eq:preference_definition}
  x_{i s} = \mathbbm{E} [ k_{i s} / k_i ],
\end{equation}
where we restrict ourselves to nodes~$i$ with non-zero degree---the value of $x_{is}$ is not well-defined when $k_i=0$.  Note that $x_{i s}$ as defined is automatically normalized so that $\sum_s x_{i s} = 1$.  Note also that the ratio $k_{i s}/k_i$ is, by definition, an unbiased estimator of $x_{i s}$, though it is not necessarily a good estimator.  In fact, as we demonstrate below, for many purposes it is highly misleading.

One way to think about Eq.~\eqref{eq:preference_definition} is to imagine creating the same network many times over and averaging over the randomness in the creation process to calculate~$x_{is}$.  Unfortunately, in the real world we normally get to observe a network only once and hence we cannot perform the average.  This is the root cause of the difficulty with estimating preferences that we mentioned above.

To proceed any further we need to know more about the nature of the fluctuations in the values of the~$k_{is}$.  If we can define a sensible model for these fluctuations then we can make progress on estimating~$x_{is}$ using the tools of statistical inference.

\subsection{Preference-based network model}
How is $k_{is}$ generated?  We could imagine that node~$i$ considers every other node in turn and connects to those in group~$s$ with some probability~$\lambda_{is}$, which measures $i$'s affinity for group~$s$.  Then the edges of the network would be Bernoulli random variables with means~$\lambda_{is}$, which in standard statistical notation would be written $A_{ij} \sim \text{Bernoulli}(\lambda_{i g_j})$, where $A_{ij}$ is an element of the adjacency matrix, having value one if there is an edge from $i$ to~$j$ and zero otherwise, and $g_j$ is the group or type label of node~$j$.

This, however, is unsatisfactory for two reasons.  First, as is often the case, it is simpler to use a Poisson rather than Bernoulli distribution: $A_{ij} \sim \text{Poisson}(\lambda_{ig_j})$.  In a sparse network where $\lambda_{is}\ll1$ the two distributions are nearly identical, but the Poisson distribution offers significant technical advantages.  Second, and more importantly, many networks have broad degree distributions that are not well captured by either the simple Bernoulli or Poisson model.  This issue can be dealt with by ``degree-correction''~\cite{KN11a,Fortunato16, Peixoto12}, which in this context involves the introduction of two additional parameters $\phi_i$~and~$\theta_i$ for each node~$i$, which respectively control the in- and out-degrees of the node.  (In an undirected network, the two would be equal~$\phi_i=\theta_i$.)  Then we let
\begin{equation}
\label{eq:Dirichlet_edge}
A_{ij} \sim \text{Poisson}\left( \frac{\theta_i \phi_j x_{i g_j}}{\Phi_{g_j}} \right)
\end{equation}
where
\begin{equation}
	\Phi_s = \sum_{i \in s} \phi_i 
\label{eq:Phi_norm}
\end{equation}
is the sum of all $\phi_i$ for nodes in group~$s$.  This definition does not completely fix the values of the parameters, since we can multiply the values of all the $\phi$ by any constant factor without affecting the~$A_{ij}$ or any other property of the model.  One can fix this by choosing a normalizing condition for the~$\phi_i$, such as requiring that they sum to~1, but this will not be necessary for any of the calculations presented here.

Note that the choice of a Poisson rather than a Bernoulli distribution in Eq.~\eqref{eq:Dirichlet_edge} implies that the network may have multiedges---there may be two or more edges running between the same pair of nodes, so that $A_{ij}>1$.  On a sparse network, however, this happens vanishingly often and multiedges can normally be neglected~\cite{KN11a}.

For a better intuition on the role of the parameters in the model, it is instructive to consider the distributions of the quantities~$k_{is}$ and~$k_i$.  Given that the $A_{ij}$ are independent Poisson random variables and that a sum of Poisson variables is itself Poisson, the distributions for~$k_{is}$ and $k_i$ are also Poisson:
\begin{equation}
k_{is} \sim \text{Poisson} \bigl( \theta_i x_{is} \bigr),
\end{equation}
and
\begin{equation}
k_i = \sum_s k_{is} \sim \text{Poisson} \left( \theta_i \right).
\end{equation}
Thus $\theta_i$ is equal to the expected (out-)degree at node~$i$, independent of the node's preferences. A simple further computation verifies that $x_{i s}$ is indeed the expected value of $k_{i s}/k_i$, consistent with the definition of preference, given in Eq.~\eqref{eq:preference_definition}.

We favor this model for the intuitive interpretation of its parameters along with the mathematical simplicity of the Poisson distribution.

\subsection{Inferring individual preferences}
Given the types of the nodes, we can now write down the probability of observing any given pattern of connections at node~$i$:
\begin{align}
\label{eq:P(A_i|x_i,g,theta,phi)}
P( \bm{A}_i | \bm{x}_{i}, g, \theta, \phi ) &= \prod_j P(A_{i j} | \bm{x}_{i},g, \theta, \phi) \nonumber \\ 
	&=  e^{-\theta_i} \prod_j \left( \frac{\theta_i \phi_j x_{i g_j}}{\Phi_{g_j}} \right)^{A_{ij}} \frac{1}{A_{ij}!},
\end{align}
where $\bm{A}_i$ denotes the $i$th row of the adjacency matrix and $\bm{x}_i$ is the vector with elements~$x_{is}$.  The probability of observing the whole network is then the product
\begin{equation}
\label{eq:P(A|x,g,theta,phi)}
 P( A | x, g, \theta, \phi ) = \prod_ i P( \bm{A}_i | \bm{x}_{i}, g, \theta, \phi ).
\end{equation}
The terms in Eq.~\eqref{eq:P(A|x,g,theta,phi)} that depend on $\theta$ and $\phi$ can be factored out from those that depend on $x$ and thus one can write
\begin{equation}
	\label{eq:P(A|x,g)}
	P(A | x, g) = \frac{1}{Z} \prod_{i, s} x_{i s}^{ k_{i s}}
\end{equation}
where $Z$ is a constant that depends on $A$ and $g$ but not~$x$.

Given both the categories and the network structure, we can use the model to infer the preferences~$\bm{x}_i$.  A~tempting approach is to use maximum-likelihood estimation.  However, maximization of Eq.~(\ref{eq:P(A|x,g)}) with the constraint that $\sum_s x_{is} = 1$ just leads back to the estimate $\hat{x}_{is} = k_{is}/k_i$.  As we now argue, if we want to learn about the distribution of preferences, these estimates may be misleading.

\begin{figure}
\centering
\includegraphics[width=0.8\linewidth]{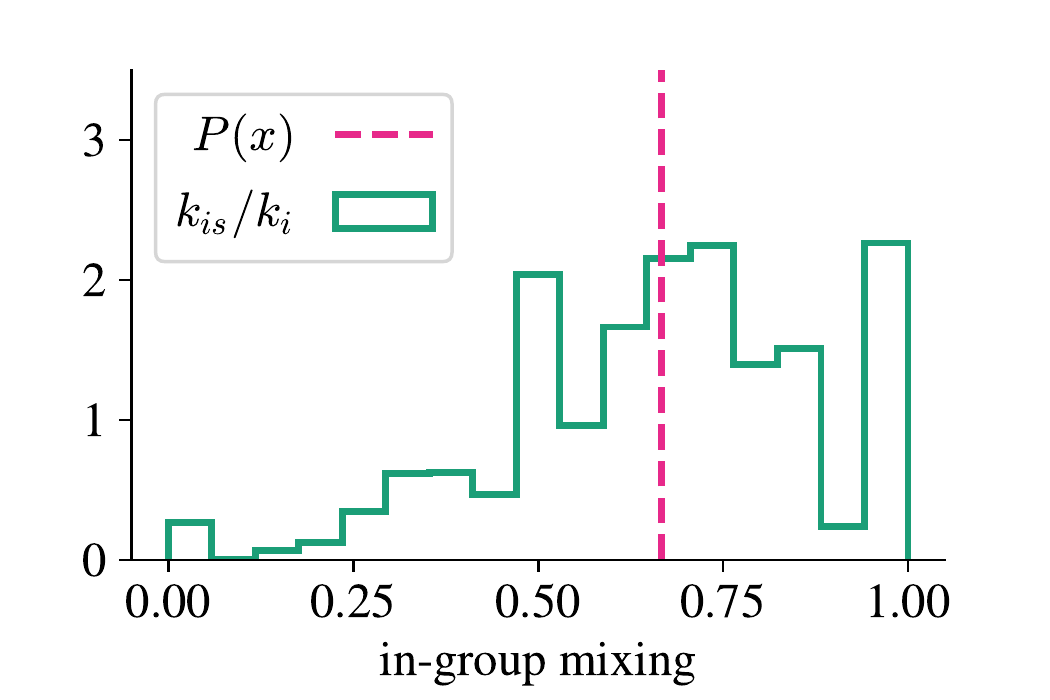}
	\caption{Histogram for $k_{i s}/ k_i$ in the model of Eq.~\eqref{eq:Dirichlet_edge} when $\theta_i=6$ and $x_{i s}=\frac23$. The dashed line is at~$\frac23$, the true value of $x_{i s}$. For an arbitrarily large network with these parameters, the dashed line is the true distribution of preferences, while the histogram correspond to the inferred distribution, if we used the naive/maximum-likelihood estimator $\hat{x}_{is}=k_{is}/k_i$. The distribution of $k_{is}/k_i$ is clearly a poor approximation for the true distribution $P(x)$.}
\label{fig:estimator_distribution}
\end{figure}

Consider, for illustrative example, the case in which all nodes in a group have the same parameter values.  (This case is equivalent to the stochastic block model.)  Even though all nodes have the same preferences, $k_{i s}/k_i$~will not be the same for every node, since it is a random variable. Worse, it will often have significant variation. Figure~\ref{fig:estimator_distribution} shows an example of this situation. 

Things are not too bad if we only want to measure the average preferences in a group: we can average over the values of $k_{is}/k_i$ for all $i$ in the group in question and the fluctuations will average out.  For anything beyond average-level behavior, however, we are not so lucky.  As demonstrated in Fig.~\ref{fig:estimator_distribution}, even something as simple as the variance of $x_{i s}$ is not straightforward to estimate from~$k_{is}/k_i$.

The root of the problem is the sparsity of the network.  When we only have a handful of connections for each node, the ratio $k_{is}/k_i$ will be broadly distributed even when all $x_{is}$ are the same.  This is not due to our networks being too small.  The amount of network data we have grows larger as the network does, but so too does the number of parameters we are estimating, and it is straightforward to show that the expected variation of the individual estimates~$k_{is}/k_i$ will not vanish even in the large size limit.

To get around this issue we need some way to accurately characterize individual preferences that does not require an extensive number of parameters.  Here we do this by inferring the underlying distribution from which the $\bm{x}_i$ are generated.  We describe this procedure in the next section.

\section{Distributions of preferences}
\label{sec:distribution_infer}
Suppose the preference variables~$x_{is}$ for nodes in group~$r$ are drawn from a distribution $P(\bm{x} |\bm{\alpha}_r)$, where $\bm{\alpha}_r$ is a set of parameters for the distribution.  If we know this distribution then we can integrate over the unobserved preferences in Eq.~\eqref{eq:P(A|x,g)} and compute
the likelihood of the network thus:
\begin{equation}
	P(A | \alpha, g) =\frac{1}{Z} \prod_i \int \Big( {\textstyle \prod_{s}} x_{s}^{k_{is}} \Big) P(\bm{x}|\bm{\alpha}_{g_i})  \>d\bm{x}.
\label{eq:likelihood}
\end{equation}
Rather than infer the individual preferences directly we can, using this likelihood, infer their distribution by fitting the parameters~$\alpha$. 

The only constraints on~$\bm{x}$ are that $x_{s}>0$ for all~$s$ and $\sum_s x_s= 1$, meaning that the vector $\bm{x}$ lies on the standard unit simplex and $P(\bm{x} | \bm{\alpha}_r)$ can be any distribution on the simplex.  Here we make the simple and common assumption that $P(\bm{x} | \bm{\alpha}_r)$ is a Dirichlet distribution~\cite{StatInference}.  For a network with $c$~groups the Dirichlet distribution takes the form
\begin{equation}
P(\bm{x} | \bm{\alpha})
  = {1\over B(\bm{\alpha})} \prod_{s=1}^c x_s^{\alpha_s-1},
\label{eq:Dirichlet_dist}
\end{equation}
where $\alpha_s>0$ for all~$s$ and $B(\bm{\alpha})$ is the multi-dimensional beta function
\begin{equation}
B(\bm{\alpha}) = {\prod_s \Gamma (\alpha_s) \over \Gamma(\alpha_0)},
\end{equation}
with $\alpha_0 = \sum_s \alpha_s$ and $\Gamma(\alpha)$ being the standard gamma function.  The Dirichlet distribution is a convenient and flexible distribution that allows us to vary the weight placed on each of the $x_s$ independently.
In the case of two groups, $c=2$, the Dirichlet distribution is equivalent to the beta distribution.  The expected value of $\bm{x}$ within the distribution is~$\bm{\alpha}/\alpha_0$, and $\alpha_0$ controls the width of the variation about that value.  In the limit of large $\alpha_0$ the variance tends to zero and the distribution of $\bm{x}$ is tightly clustered around the mean.  Conversely, as $\alpha_0$ tends to zero almost all the probability density is in the corners of the simplex, as far away as possible from the mean.

We allow each group or type~$s$ to have a different distribution of preferences and hence a different set of Dirichlet parameters~$\bm{\alpha}_s$, so that the prior on~$\bm{x}_i$ is
\begin{equation}
\bm{x}_i \sim \text{Dirichlet}(\bm{\alpha}_{g_i}).
\end{equation}
This is a natural choice: one can well imagine, for instance, that the men and women within a population have different preferences for male and female friends.

With this choice we can now complete the integrals in Eq.~\eqref{eq:likelihood} and we find that
\begin{equation}
	\label{eq:P(A|alpha,g)}
	P(A|\alpha,g) = \frac{1}{Z} \prod_i \frac{B(\bm{\alpha}_{g_i} + \bm{k}_i)}{B(\bm{\alpha}_{g_i})},
\end{equation}
where $\bm{k}_i$ is the vector with elements~$k_{is}$.  Estimates for $\alpha$ can now be obtained by maximizing this likelihood.

Under certain circumstances, Eq.~\eqref{eq:P(A|alpha,g)} may lack a well-defined maximum. To deal with this one can add a regularization term. The full details are given in Appendix~\ref{appendix:a}, but the end result is that one determines the estimated value $\hat{y}_{rs}$ by maximizing
\begin{align}
	\label{eq:reg_likelihood}
	L(y) =  \sum_{i}  \bigl[ \ln  & B(e^{\bm{y}_{g_i}} + \bm{k}_i)  - \ln  B(e^{\bm{y}_{g_i}}) \bigr] - \lambda \sum_{r,s} y_{rs}^2,
\end{align}
where $\lambda$ is a small positive constant, and our estimate of $\alpha_{rs}$ is given by $\hat{\alpha}_{rs} = \exp \hat{y}_{rs}$. From a Bayesian perspective the quadratic regularization term is equivalent to a log-normal prior on~$\alpha_{rs}$.

Our reasoning up to this point can be summarized as follows.  When we try to directly infer node preferences we find that the distribution of our estimates does not in general resemble the true underlying distribution, even for arbitrarily large networks.  In contrast, maximization of Eq.~\eqref{eq:reg_likelihood} should give accurate estimates of~$\alpha$, at least for large networks, and to the extent that the underlying distribution can be well fit by the hypothesized Dirichlet distribution, these parameters will describe the shape of that distribution.  Thus, it is now possible to infer preference distributions accurately so long as the network is sufficiently large.

In the real world we don't have arbitrarily large networks and so a different source of error could arise: the inability to make accurate estimates of~$\alpha$ because our data are limited.  One way to get around this problem is to take a Bayesian approach.

Bayes' theorem states
\begin{equation}
	\label{eq:Bayestheorem_alpha}
	P(\alpha | A, g) = \frac{ P(A|\alpha, g) P(\alpha)}{P(A|g)}.
\end{equation}
The distribution $P(\alpha)$ is the prior distribution for the parameters, which we have to choose.  Since the regularization term introduced in Eq.~\eqref{eq:reg_likelihood} is equivalent to a log-normal prior for~$\alpha_{rs}$, we propose using this form as a prior.  More details are given in Appendix~\ref{appendix:a}.

A posterior distribution on~$\alpha$ as above allows us to make estimates of quantities of interest without having to estimate~$\alpha$ itself---we can average over it instead.  In the next section we define some useful metrics that can be evaluated within the posterior distribution, and can thus be inferred in a parameter-free way.

\section{Measures of assortativity and variation of preferences}
\label{sec:measures}
In the previous section we described a procedure for inferring preference distributions in networks.  The full multi-dimensional distribution, however, is difficult to interpret physically, so simple summary statistics are also useful.  In this section we propose two specific measures that quantify the average assortativity in the network and the variation of preferences around that average.

Assortative mixing occurs when nodes have a preference for connecting to others of the same type.  A natural measure of assortativity is the expected value of the in-group preference parameters.  As discussed in Section~\ref{sec:distribution_infer}, the expected value of the preference parameter~$x_{is}$ describing the preference of a node~$i$ in group~$r$ for connections to group~$s$ is $\alpha_{rs}/\alpha_{r0}$ where $\alpha_{r0} = \sum_s \alpha_{rs}$.  The expected in-group preference of nodes in group~$r$---their preference to connect to other members of the same group---is then equal to $\alpha_{rr}/\alpha_{r0}$, and the average in-group preference over all nodes in all groups is
\begin{equation}
a = \sum_r p_r {\alpha_{rr} \over \alpha_{r0}},
\label{eq:assortativity}
\end{equation}
where $p_r$ is the fraction of nodes that fall in group~$r$.

In a perfectly assortative network all nodes connect only to their own group and $a=1$, while in a perfectly disassortative network $a=0$.  For most real-world networks we expect the value to lie between these extremes, with higher values indicating more assortativity.  A natural question to ask is what kinds of values do we expect to see?  What constitutes a ``high'' value of~$a$?  One way to answer this question is to calculate the expected value within a null model.

A suitable null model in this case is one in which nodes are connected according to their expected degrees,
\begin{equation}
A_{ij} \sim \text{Poisson} \biggl( \frac{k_i k^{\textrm{in}}_j }{m} \biggr),
\end{equation}
where $k_i$ denotes the out-degree of node~$i$, as previously, $k^{\textrm{in}}_i$ denotes the in-degree, and $m=\sum_i k_i$ is the expected number of edges in the network.  This is in essence just a directed version of the standard random network model in which we fix the expected degrees of all nodes, sometimes called the Chung--Lu model after two of the first researchers to examine its properties~\cite{CL02a}.

Applying the definition of preference from Eq.~\eqref{eq:preference_definition}, all nodes in group~$s$ have the same preference in this null model, $x_{is}=K_s/m$, where $K_s=\sum_{i \in s} k^{\textrm{in}}_i $. Hence in this model the average in-group preference is
\begin{equation}
	a_{\textrm{null}} = \sum_r p_r \frac{K_r}{m}.
\end{equation}

The difference between the observed value of~$a$, Eq.~\eqref{eq:assortativity}, and the expected value within the null model is then
\begin{equation}
a - a_{\textrm{null}} = \sum_r p_r \biggl( \frac{\alpha_{rr}}{\alpha_{r0}} - \frac{K_r}{m} \biggr).
\end{equation}
When this quantity is greater than zero the preferences are more assortative than we would expect by chance.  When it is less than zero the preferences are less assortative (or more disassortative) than expected.  If we wish, we can normalize the difference so that it takes a maximum value of~1 at perfect assortativity, and thus define a preference assortativity coefficient
\begin{equation}
\label{eq:assortativity-coef}
	R(\alpha) = \frac{\sum_r p_r ( \alpha_{rr}/\alpha_{r0} - K_r/m )}{\sum_r p_r ( 1 - K_r/m )}.
\end{equation}
The range of allowed values is $R~\in~\left[R_{min},1\right]$, where in general $R_{min} \neq -1$ and depends on the network in question.  (A similar behavior is seen for the conventional coefficient of assortativity defined in~\cite{Newman03c}, which is essentially a Pearson correlation.)

In order to estimate the value of~$R$, we need first to estimate the $\alpha$ parameters.  As discussed in the previous section, we could do this by maximizing the likelihood of Eq.~\eqref{eq:P(A|alpha,g)}, but this may give poor estimates in cases, such as smaller networks, where the amount of available data is limited.  An alternative approach is to compute the expected value of $R$ in the posterior distribution of Eq.~\eqref{eq:Bayestheorem_alpha}, thus:
\begin{equation}
R = \int  R( \alpha  ) P(\alpha|A,g) \>d\alpha.
\end{equation}
We can also compute the standard deviation of $R$ in the posterior which makes it easy to state estimates with error bars. More details on this calculation are given in Appendix~\ref{appendix:b}.

The quantity $R$, however, only measures traditional assortativity.  As we have said, our main purpose is to examine variation of individual preferences about group means.  The variance of a Dirichlet distribution can be quantified by the mean-squared distance from its average.  In group~$r$ this is
\begin{equation}
\sigma^2_r = \mathbbm{E} \bigl[ (\bm{x} - \langle \bm{x} \rangle_r)^2 \bigr]
  = \frac{1 - \sum_s ( \alpha_{r s}/\alpha_{r 0} )^2 }{\alpha_{r 0} + 1}.
\end{equation}
As discussed in Section~\ref{sec:distribution_infer}, the maximum value of the variance occurs when $\alpha_{r0}\to0$, which gives $\sigma^2_r = 1-\sum_s ( \alpha_{r s}/\alpha_{r 0} )^2$.  One can divide by this maximum to give a normalized variance
\begin{equation}
V_r = {\sigma_r^2\over1 - \sum_s ( \alpha_{rs}/\alpha_{r0} )^2}
    = {1\over\alpha_{r0}+1},
\end{equation}
which lies between zero and one and also has the nice property of being independent of the mean.  Then we can define an overall normalized variance coefficient by
\begin{equation}
	V(\alpha) = \sum_r p_r V_r = \sum_r \frac{p_r}{\alpha_{r0} + 1},
\end{equation}
which also lies between zero and one. $V$ can be estimated in the same way as $R$ by averaging its value in the posterior distribution. See Appendix~\ref{appendix:b} for further details of this calculation. 

The quantity~$V$ represents the normalized mean-square distance between the preferences and their group means, averaged over all groups.  When $V$ is close to zero every node in every group has preferences close to the group mean.  If preferences are homogeneous in this way then the network is well described by the group average mixing parameters and individuals' preferences are well described by simply stating which group they belong to.  Such a finding could be informative for instance in a social network: it would tell us a lot about a population if we found that their preferences were entirely determined by, say, gender or race.

At the other extreme, when $V$ approaches one, node preferences are as far away from the group mean as possible, and nodes, even within the same group, are very unlike each other in their preferences.  In this scenario mixing is poorly described by average rates, since virtually no nodes behave according to the average for their group.  

\begin{table}
\setlength{\tabcolsep}{4pt}
\centering
\begin{tabular}{l|cc}
Network & $\ R$ & $V$ \\
\hline
	College football~\cite{NG04}                    & $\ \ 0.60 \pm 0.015$ & $0.01 \pm 0.004$ \\
Karate club~\cite{Zachary77}                    & $\ \ 0.72 \pm 0.063$ & $0.07 \pm 0.059$ \\
Political books                                 & $\ \ 0.72 \pm 0.028$ & $0.12 \pm 0.034$ \\
Political blogs~\cite{AG05}                     & $\ \ 0.80 \pm 0.010$ & $0.15 \pm 0.012$ \\
High school race or ethnicity~\cite{AddHealth}     & $\ \ 0.55 \pm 0.012$ & $0.16 \pm 0.011$ \\
Provisional IRA affiliation~\cite{gill2014lethal}                                 & $\ \ 0.62\pm0.025$ & $0.22 \pm 0.026$ \\
Word adjacency~\cite{kucera1979standard}        & $-0.27 \pm 0.018 $ & $0.30 \pm 0.024$ \\
\end{tabular}
\caption{Estimates of normalized preference assortativity $R$ and preference variance $V$ for a selection of networks with known group assignments. Results are computed from the posterior distribution and stated as $\mu_{X} \pm \sigma_{X}$.  Numbers in brackets indicate references for each network, except for the network of political books, which was compiled by Valdis Krebs and is currently unpublished.}
\label{table:examples}
\end{table}

\begin{figure*}
\centering
\includegraphics[width=1\linewidth]{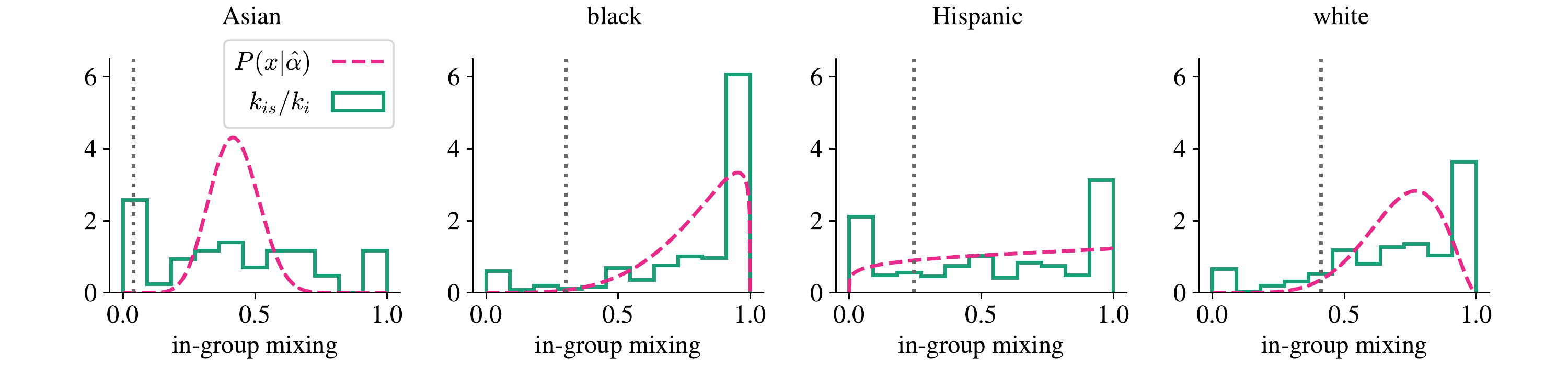}
	\caption{Friendship preferences by race or ethnicity in a US high school.  We show separate results for Asian, black, Hispanic, and white students.  For each race or ethnicity the histogram (in green) shows the observed distribution of~$k_{ig_i}/k_i$, the naive estimate of within-group preference.  The red dashed line is the inferred preference distribution from a point estimate of $\alpha$, found by maximizing Eq.~\eqref{eq:reg_likelihood}.  The gray vertical line is where the average preference would be, in the absence of assortativity.}
\label{fig:race_high_school} 
\end{figure*}

\section{Examples}
Table~\ref{table:examples} shows results for the preference assortativity and variance measures, $R$~and~$V$, for a selection of previously studied networks with known group assignments, listed in order of increasing variance.  As the table shows, all of the networks are highly assortative by our measure, except for the word adjacency network, which is disassortative.

The normalized variances~$V$ take a range of values from zero up to 0.3.  Recall that low normalized variance indicates a network in which the members of a group have similar preferences; high variance indicates that they have widely varying preferences.  Thus, for instance, the ``karate club'' network, which is a social network of university students, appears to have no significant variance, meaning it shows traditional community structure in which the members of a community are roughly alike in their preferences.  The network of high school students, on the other hand, which one might expect to be similar, shows higher variance.  We discuss the high school and word networks in more detail below.

\subsection{High school friendships and ethnicity}
\label{section:highschool_example}
The network denoted ``High school race or ethnicity'' in Table~\ref{table:examples} is a network of self-reported friendships between students in a US high school, taken from the National Longitudinal Study of Adolescent to Adult Health~\cite{AddHealth} (commonly known as the ``Add Health'' study).  The node labels in this case represent the (self-identified) ethnicities of the students, which take values ``Asian,'' ``black,'' ``Hispanic,'' ``white,'' ``other,'' and ``missing.''  In our analysis we discard the ``other'' and ``missing'' categories and focus on the remaining four.  The particular school we look at is chosen for its diverse racial and ethnic composition.

The value of $R=0.55 \pm 0.012$ for this network indicates that the school is strongly assortative by race, meaning that students had more within-group friendships than would be expected by chance.  However, the groups also display differences in the inferred distributions of their preferences, which are plotted in Fig.~\ref{fig:race_high_school}.  Hispanic students, for instance, show a larger range of preferences than others.  Note that this doesn't necessarily imply that Hispanic students individually have diverse friendship groups---some of them do, but others show a strong preference for having mainly Hispanic friends, or for having few.

Also shown in Fig.~\ref{fig:race_high_school} are histograms of the naive preference estimates~$k_{is}/k_i$, which look quite different from the inferred distributions.  This discrepancy is expected: as discussed in Section~\ref{sec:preferences}, the distribution of naive estimates is an unreliable indicator of the true preference distribution.

\begin{figure*}
\centering
\includegraphics[width=0.75\linewidth]{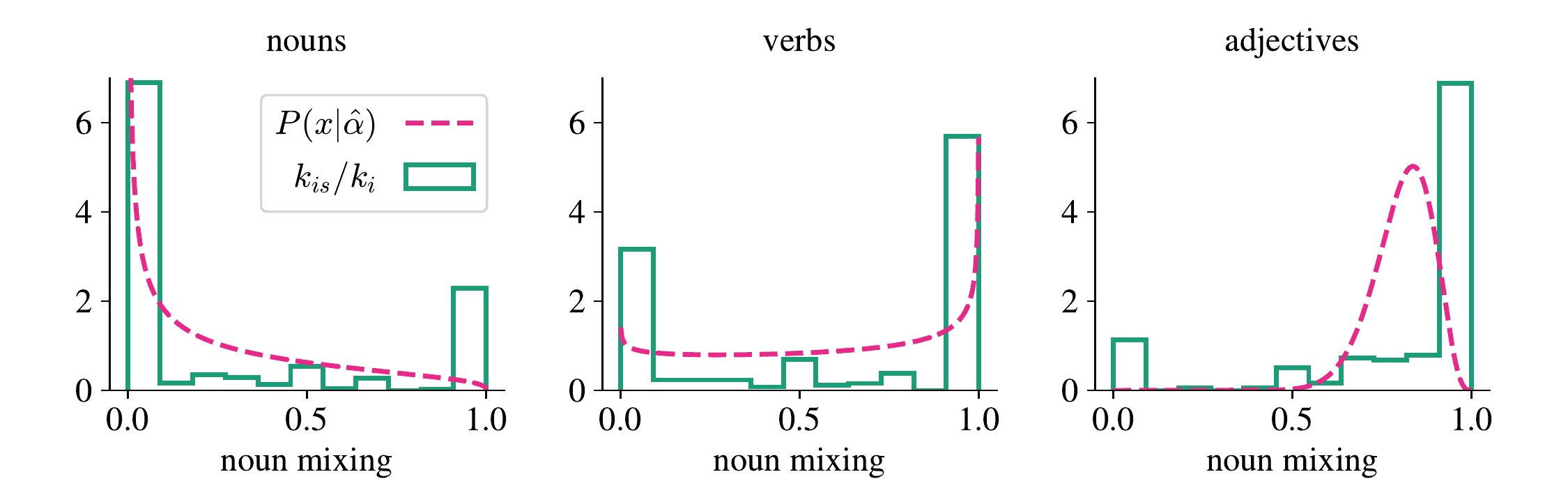}

	\caption{Preferences of different parts of speech to be followed by nouns. Each word is followed by a noun some proportion of the time, and this proportion is different for different words. For each type of word the histogram (in green) shows the observed distribution of~$k_{i, \text{noun}}/k_i$, the naive estimate of noun preference.   The red dashed line is the inferred preference distribution from a point estimate of $\alpha$, found by maximizing Eq.~\eqref{eq:reg_likelihood}. The three plots represent the distributions for nouns, verbs, and adjectives from the fiction portion of the Brown corpus of English text~\cite{kucera1979standard}.}
\label{fig:word_adjacency}
\end{figure*}

\subsection{Word adjacencies}
\label{section:words}
The \textit{Brown corpus} is a widely used data set consisting of samples of written English text compiled by researchers at Brown University in the 1960s~\cite{kucera1979standard}.  Words in the data set are labeled with their part of speech---noun, adjective, verb, etc.  Working from the fiction text contained in the corpus, we create a directed word adjacency network in which nodes represent words (limited to nouns, adjectives, and verbs) and there is a directed edge from word~$i$ to word~$j$ if word~$i$ is followed by word~$j$ at any point in the text.

Figure~\ref{fig:word_adjacency} shows the inferred distributions of preferences within this network for nouns, verbs, and adjectives to be followed by nouns.  For example, since adjectives normally come before nouns in English we would expect adjectives to have a preference for being followed by nouns.  And indeed this is what we see---the red curve in the third panel of Fig.~\ref{fig:word_adjacency} shows that most adjectives have a high preference for being followed by nouns.  Nouns, on the other hand, aren't usually followed by other nouns, although they can be: the distribution (shown in the first panel of the figure) takes its most likely value around a preference of zero, but is spread across the whole range and there is still a relatively large density around preference~1, which is to say that some nouns strongly prefer to be followed by other nouns.  Classic examples are titles such as ``Mr.''\ and ``Mrs.,'' which are almost always followed by proper nouns.  Likewise, although most verbs prefer to be followed by nouns, there are a handful that have a strong preference to be followed by another verb.  These are typically auxiliary verbs, such as ``has'' and ``was'', in sentences like ``He was sleeping.''

\section{Conclusions}
In this paper we have considered the problem of characterizing mixing patterns in networks.  Average mixing patterns have a long history of study and can be quantified using standard methods, but anything beyond the average requires additional machinery for its description.  We analyze within-group variation in mixing using a model of individual preferences in networks, showing how to fit the model to data using Bayesian methods.  The parameters of the fit have simple interpretations and we use them to define coefficients that quantify the average assortativity and variation of preferences.  The method is computationally efficient, with running time growing linearly in the size of the data set, which puts applications to large networks within reach.

We have given applications of our methods to a range of social and information networks.  We find that some, though not all, of these networks do display significant within-group variation in their mixing patterns, and that where such variation is present the mixing is not well described by traditional community structure.  Even when there is little or no variation in preferences the analysis is still informative, since it implies that preferences are well described solely by which group a node belongs to.

A limitation of our approach is the assumption that the preferences are drawn from a Dirichlet distribution, which rules out multimodal distributions for example.  One natural avenue of extension for the approach would be to experiment with other choices of distribution.  Instead of a single Dirichlet distribution, for example, one could use a mixture (i.e.,~a linear combination) of two or more.  This would allow us to model more complex behaviors, at the expense of a more complicated fitting procedure.  We have also here considered only the case in which the label or group membership of every node is known.  One could generalize our methods to deal with cases in which all or some of the data are unknown, but we leave these developments for future work.

\begin{acknowledgments}
The authors thank Elizabeth Bruch, Alec Kirkley, and Maria Riolo for useful conversations.  This work was funded in part by the US National Science Foundation under grant DMS--1710848.
\end{acknowledgments}

\begin{center}
\rule{4cm}{0.5pt}
\end{center}

\appendix

\section{Point estimates for $\bm{\alpha}$}
\label{appendix:a}
The maximum likelihood estimate for $\bm{\alpha}_r$ is given by the location of the maximum of
\begin{align}
	\label{eq:L_r(alpha)}
	L_r(\bm{\alpha}_r) &= \sum_{i \in r} \bigl[ \ln  B(\bm{\alpha}_{r} + \bm{k}_i)  - \ln  B(\bm{\alpha}_{r}) \bigr].
\end{align}
Here $\ln B(\bm{x})$ is the log of the multivariate beta function,
\begin{equation}
\ln B(\bm{x}) = - \ln \Gamma(x_0) + \sum_s \ln \Gamma(x_s),
\end{equation}
with $x_0 = \sum_s x_s$.
Both the Jacobian and Hessian of $L_r$ are straightforward to compute, so in principle one could perform the maximization using optimizers such as Newton's method that require second derivatives. 

There are however some technical complications with direct maximization of~\eqref{eq:L_r(alpha)}.  First, one must impose the constraint $\alpha_{rs}>0$, which can be done by re-parameterizing with $y_{rs} = \ln \alpha_{rs}$ and writing
\begin{align}
	L_r(\bm{y}_r) &= \sum_{i \in r} \bigl[ \ln  B(e^{\bm{y}_{r}} + \bm{k}_i)  - \ln  B(e^{\bm{y}_{r}}) \bigr].
\end{align}
An unconstrained maximization with respect to $\bm{y}_r$ then achieves the desired goal.

Second, and more important, under some circumstances the maximum is not guaranteed to exist and $L_r$ can increase as $y_{rs}\to \pm \infty$.  For a well-defined estimate we must insist on a maximum at a finite value of~$y_{rs}$.  A simple way to do this is to add a quadratic regularization term to the likelihood thus:
\begin{align}
\label{eq:regularized_Likelihood}
L_r(\bm{y}_r) &=  \sum_{i \in r} \bigl[ \ln  B(e^{\bm{y}_{r}} + \bm{k}_i)  - \ln  B(e^{\bm{y}_{r}}) \bigr]  - \lambda \sum_s y_{rs}^2,
\end{align}
where $\lambda$ is a small positive constant.

From a Bayesian perspective this quadratic regularization corresponds to placing a normal prior on~$y_{rs}$ with mean zero and variance $(2 \lambda)^{-1}$, or equivalently a log-normal prior on~$\alpha_{rs}$.  As $\lambda \to 0$ the prior on $y_{rs}$ becomes uniform, so any small fixed value of $\lambda$ should give acceptable results.  We use $\lambda=2^{-7}$, equivalent to $\sigma=8$, which implies that $\alpha_{rs}$ falls roughly between the $3\sigma$ bounds $10^{-10}$ and~$10^{10}$.

To find the maximum of Eq.~\eqref{eq:regularized_Likelihood} one can use any numerical optimization technique.  For techniques that make use of the Jacobian and/or Hessian, the Jacobian is given by
\begin{align}
\label{eq:jacobian}
\frac{\partial L_r}{\partial y_{r s}} = e^{y_{rs}} \sum_{i \in r} \bigl[
	&\psi(e^{y_{rs}} + k_{is})  - \psi \bigl( \textstyle\sum_t e^{y_{rt}} + k_{i} \bigr) \nonumber \\
	& - \psi(e^{y_{rs}}) + \psi \bigl(\textstyle\sum_t e^{y_{rt}} \bigr) \bigr] - 2 \lambda y_{rs},
\end{align}
where $\psi(x) = \Gamma'(x)/\Gamma(x)$ is the so-called digamma function. The Hessian is given by
\begin{align}
	\frac{\partial^{2} L_r}{\partial y_{r s}^2} &=  e^{y_{rs}} \frac{\partial L_r}{\partial y_{r s}}  + e^{ 2 y_{rs}} \sum_{i \in r} \bigl[
		\psi' (e^{y_{rs}} + k_{is})  \nonumber \\
		&- \psi' \bigl( \textstyle\sum_t e^{y_{rt}} + k_{i} \bigr) - \psi'(e^{y_{rs}}) + \psi' \bigl(\textstyle\sum_t e^{y_{rt}} \bigr) \bigr] - 2 \lambda , \nonumber
\end{align}
\begin{align}
	\label{eq:hessian}
	\frac{\partial^{2} L_r}{\partial y_{r s} \partial y_{r t}} &=  e^{y_{rs}+y_{rt}} \sum_{i \in r} &\bigl[
		 \psi' \bigl(\textstyle\sum_t e^{y_{rt}} \bigr)  - \psi' \bigl( \textstyle\sum_t e^{y_{rt}} + k_{i} \bigr) \bigr],
\end{align}
where $\psi'(x)$ is the trigamma function.

\section{Bayesian estimates for $R$ and $V$}
\label{appendix:b}
To compute an estimate of any quantity that depends on~$\alpha$, we can average its value over the posterior distribution.  For any function~$f(\alpha)$ the average is given by
\begin{equation}
\langle f \rangle = \int  f(\alpha) P(\alpha|A,g) \>d\alpha,
\end{equation}
which can also be written
\begin{equation}
\langle f \rangle = \frac{ \int f(y) \exp\bigl[ {\textstyle\sum_r} L_r(\bm{y}_r) \bigr] \>dy} { \int \exp\bigl[ \textstyle{\sum_r} L_r(\bm{y}_r) \bigr] \>dy},
\end{equation}
where $y_{r s} = \ln \alpha_{r s}$ and $L_r(\bm{y}_r)$ is defined by Eq.~\eqref{eq:regularized_Likelihood}. 

Both $R$ and $V$, as we have defined them, are averages over the groups, $R = \sum_r p_r R_r$ and $V = \sum_r p_r V_r$. For any such function we can compute the averages for the individual groups separately
\begin{equation}
\label{eq:F_average}
\langle F \rangle = \sum_r p_r \langle F_r \rangle = \sum_r p_r \frac{ \int F_r(\bm{y}) \exp\bigl[ L_r(\bm{y}) \bigr]  \>d\bm{y}}{ \int \exp \bigl[ L_r(\bm{y}) \bigr] \>d\bm{y}}.
\end{equation}
Integrals of this form can be approximated using Laplace's method, which in this case gives
\begin{equation}
\label{eq:Laplace_approx}
\langle F_r \rangle \simeq \sqrt{\frac{\det \bm{\Sigma}_r^{\bm{*}}}{\det \bm{\Sigma}_r}}\, \exp \bigl[ L_r^{*}(\bm{\hat{y}}_r^{\bm{*}}) - L_r(\bm{\hat{y}}_r) \bigr],
\end{equation}
where
\begin{equation}
	L_r^{*}(\bm{y}) = L_r(\bm{y}) + \ln F_r(\bm{y}),
\end{equation}
\begin{align}
	\label{eq:y_dagger}
	\bm{\hat{y}}_r &= \argmax_{\bm{y}} \left\{ L_r(\bm{y}) \right\}, \\
	\bm{\hat{y}}_r^{\bm{*}} &= \argmax_{\bm{y}} \left\{ L_r^{*}(\bm{y}) \right\}, 
\end{align}
and $\bm{\Sigma}_r^{\bm{*}}$ and $\bm{\Sigma}_r$ are minus the inverse of the Hessians of $L_r^{*}$ and $L_r$ at $\bm{\hat{y}}_r^{\bm{*}}$ and $\bm{\hat{y}}_r$. 
In this ratio form some errors cancel and Laplace's approximation has only an $O(n^{-2})$ error~\cite{TierneyKadane1986}. 

Estimates for $R$ and $V$ can now be computed from Eqs.~\eqref{eq:F_average} and~\eqref{eq:Laplace_approx} with
\begin{align}
F^{(R)} &= \sum_r p_r \frac{e^{y_{rr}}}{\sum_s e^{y_{rs}}}, \\
F^{(V)} &= \sum_r p_r \frac{1}{1+ \sum_s e^{y_{rs}}}.
\end{align}
The values of $\bm{\hat{y}}_r$ and $\bm{\hat{y}}_r^{\bm{*}}$ along with the Hessians can be computed from Eqs.~\eqref{eq:jacobian} and~\eqref{eq:hessian}.  Error estimates can also be computed from estimates of $R^2$ and~$V^2$.

Software to compute estimates of $R$ and $V$ is available in Ref.~\cite{individual_mixing_github}.

\end{document}